\documentclass{cernrep}

\usepackage{subcaption} 
\usepackage[T1]{fontenc}
\usepackage[bookmarks, colorlinks=true, linktoc=page, pdftex, linkcolor=black, citecolor=black, urlcolor=blue]{hyperref}

\usepackage{fancyhdr}
\fancyhfoffset{4 mm}
\fancypagestyle{ARTTITLE}{%
\fancyhf{} 
\lhead{\small{CERN Accelerator School Proceedings ---
{\it Advanced Accelerator Physics} ---  Spa,  Belgium, 2024}}
\lfoot{\hspace{3mm} Available online at \url{https://cas.web.cern.ch/previous-schools}}
\rfoot{\thepage\hspace*{3mm}}
 
}

\usepackage[labelfont=bf]{caption}
\usepackage{varwidth}
\usepackage{xcolor}
\begin{document}

\title{Energy Recovery LINACs - an Overview}
\author{M. Arnold}
\institute{Institut für Kernphysik, Fachbereich Physik, Technische Universität Darmstadt, Darmstadt, Germany}

\begin{abstract}
    The seminar on energy recovery linacs (ERLs) is giving an overview of the~field: How does an ERL work? What have been important milestones in ERL history? What are the reasons to use an ERL instead of a conventional accelerator? As examples of the landscape of machines, ranging from ancient ERLs up to future projects, this chapter will give results of the runs from CBETA (USA) and \mbox{S-DALINAC} (Germany). The two facilities bERLinPro/SEALab (Germany) and MESA (Germany) will belong to the next ERLs to be in operation and will be introduced briefly. The way to future ERLs will also be addressed.
\end{abstract}


\maketitle
\thispagestyle{ARTTITLE}
\section{Introduction}
\label{sec:ERL_introduction}

    ERLs are an emerging and very exciting type of accelerator. They combine the high beam quality of linear accelerators with high average beam currents (e.g. JLabFEL with 10~mA and even higher currents for future ERLs, see also Fig.~\ref{fig:overview_ERLs}) of circular accelerators while aiming at a sustainable operation. An ERL is an electron accelerator, thus this chapter is only dealing with this type of particles. Although they are ultra-relativistic at an energy of some MeV, phase slippage is an issue and has to be taken into account in complex simulations. It is also typically using superconducting cavities for the acceleration. Here, the~radio-frequency (RF) power is mainly going directly to the beam, in contrast to a normal conducting acceleration where most of the RF power is heating up the cavity itself.

    This seminar was also given in a slightly shorter version during JUAS. This text is based on the~corresponding JUAS proceedings \cite{ERL:JUAS_Proceeding_Arnold}, extended by additional content of the CAS lectures.

    \subsection{Motivation}
    \label{subsec:ERL_motivation}
    
        The energy consumption of an accelerator facility is huge, especially for larger facilities such as the LHC. Using resources sustainably, in particular energy during operation, is getting more and more important nowadays. At the same time research activities ask for even higher center-of-mass energies and beam currents to allow for experiments with high luminosities. Looking closer at the power balance needed for the operation of a superconducting accelerator, the basic structures are:
        
        \begin{itemize}
            \item Particle production (partly operated at 2~K)
            \item Beam preparation (partly operated at 2~K)
            \item Beam acceleration (operated at 2~K)
            \item Experiment
            \item Beam dump.
        \end{itemize}
        
        \noindent The corresponding technologies have to be pushed to save energy during the operation. Examples are the~operation at 4~K by using advanced materials for the cavities increasing average field gradients and/or duty cycle or the re-use of RF power during the acceleration process. The LHeC is aiming at a~beam power of 1~GW - this would correspond to the electric power produced by a nuclear power plant or 200 to 500 on-shore wind turbines. This is obviously neither possible nor acceptable for society. Thus the hard limit of 100~MW was defined as wall-plug power consumed for the electron beam, including a~(virtual) beam power of 1~GW \cite{ERL:LHeC2012}. The big question is now: How is it possible to generate a beam of a~(virtual) beam power of 1~GW with only 100~MW when requiring ''linac-quality'' beam properties? The~solution: The electron machine for the LHeC has to be an ERL.

    \subsection{Operating Principle}
    \label{subsec:ERL_principle}
    
\begin{figure}[h]
          \begin{center}
            \includegraphics[width=0.75\textwidth]{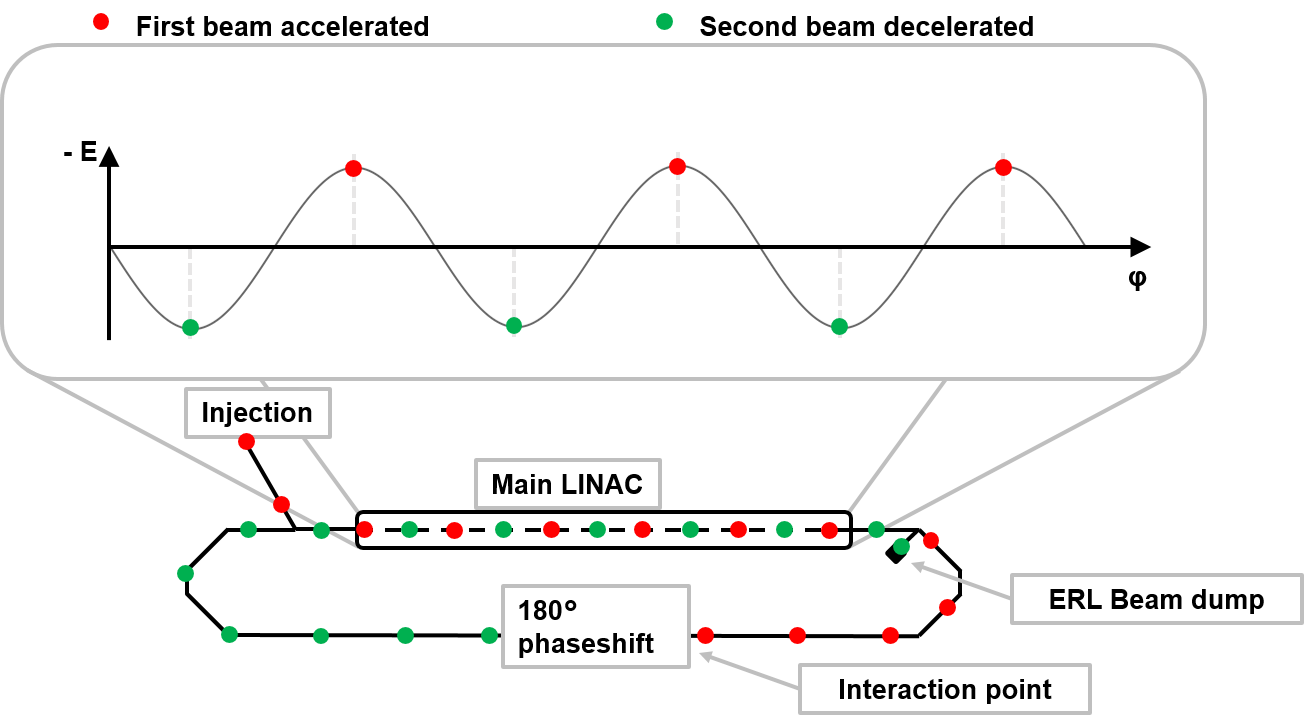}
          \end{center}
        \caption{The electron bunches from the injector (red) are accelerated and recirculated. While travelling through the recirculating beamline, a phase shift of 180° is applied, and the beam is available for an~interaction. The electrons re-enter the main accelerator on the decelerating phase (green). They are dumped at injection energy after the decelerating process.}
        \label{fig:ERL_principle}
        \end{figure}

        Figure~\ref{fig:ERL_principle} describes the basic principle of an ERL. An electron bunch (red) is coming from an injector accelerator to a main accelerator. Its timing is adjusted so that the bunches are accelerated on-crest of the~RF field of the main accelerator. Downstream of the main linac, the bunches are bent and recirculated. The bunches can be used for an experiment (see also Section~\ref{subsec:ERL_applications}) while passing the recirculation loop. At the same time a phase shift of 180° is applied. The bunches (green) re-enter the main accelerator now on the decelerating phase. Their kinetic energy is given back to the RF field while the bunches are decelerated to injection energy at the same time. The bunches are dumped at injection energy, after the~deceleration process.

    \subsection{History and Milestones}
    \label{subsec:ERL_history-milestones}

        The ERL technology had its origin in the year 1965 with the postulation of different set-ups to enable the~recovery of the beams' kinetic energy by Maury Tigner \cite{ERL:Tigner}. Around 20 years later a group at Stanford managed to recover the energy while decelerating in the same cavity as accelerating (''same-cell energy recovery'' \cite{ERL:Stanford}). This ERL operation was still without user operation. It took the accelerator community another around 14 years to have a running ERL during user operation. JLab managed to operate an~FEL while recovering the energy \cite{ERL:JLab-FEL, ERL:JLab-FEL2}. The JLab FEL is the first and so far only ERL that operated at a~beam power of more than 1~MW. Up to now all ERL efforts concentrated on a single-turn operation. In 2008 the~normal conducting ERL at BINP in Novosibirsk demonstrated the worlds' first multi-turn ERL operation \cite{ERL:BINP}. The R\&D for superconducting multi-turn ERLs took another decade: In December 2019 CBETA operated as a four-turn ERL as first superconducting multi-turn ERL, albeit only with a~small beam current and huge losses at the end of the beamline \cite{ERL:CBETA-4-turn}. In August 2021 the \mbox{S-DALINAC} achieved the first performant superconducting multi-turn ERL run \cite{ERL:S-DALINAC-2-turn}. The next ERL projects are in progress and will add new milestones on the way to future ERLs (see also Section~\ref{sec:ERL_aroundWorld}). For more details see Refs.~\cite{ERL:ERL-Roadmap, ERL:ERL-review_Hutton}.

    \subsection{Advantages of an ERL}
    \label{subsec:ERL_advantages}

        The operation of an ERL has advantages in contrast to other accelerators. As already mentioned in Section~\ref{subsec:ERL_motivation}, an ERL is the only accelerator design to achieve high beam powers while maintaining the~power consumption at a reasonable level and preserving the beam quality of a linac. This makes ERLs to the~perfect electron machine for high-luminosity colliders. Even the dumping of the beam downstream of the~collision and deceleration saves power. The beam is stopped at injection energy and thus has much less power than the high-energy beam in case of a conventional accelerator. This saves energy in cooling the beam dump or even allows for an air cooling. In case of an injection energy below 10~MeV, most of the activation of the beam dump and surrounding components can be prevented (neutron separation threshold for: \textsuperscript{27}Al=13.1~MeV, \textsuperscript{52}Cr=12~MeV, \textsuperscript{56}Fe=11.2~MeV, \textsuperscript{55}Mn=10.2~MeV, \textsuperscript{59}Ni=9~MeV, \textsuperscript{207}Pb=6.7~MeV \cite{ERL:NeutronSepThres}). Besides the operational advantages, the field of ERL contains many exciting R\&D topics, e.g. the enhancement of multi-turn operation (higher beam powers with high recovery rates, different lattice designs,...), the complex area of effects due to high bunch charges (beam break-up, microbunching,...) and also addresses accelerator R\&D in general (electron source, enhancement of SRF materials/cavities,...).

    \subsection{Possible Applications}
    \label{subsec:ERL_applications}

        ERLs can be used for all applications that have a negligible influence on the electron beam. If the experiment affected the beam too much, e.g. in a fixed-target experiment with a solid target foil, the beam could not be transported properly back to the main linac without losing too many particles. Such a loss would not only reduce transmission and, hence, efficiency, but also endanger accelerator operations due to energy deposition in the cryogenic section and/or activation. The following exemplary interactions can profit from an ERL:

        \begin{itemize}
            \item Free-electron laser (FEL): The electrons are bent on a sine-shaped trajectory in a dedicated magnetic device (undulator). The emitted photons can achieve a very high brilliance at wavelengths down to 1~\AA \hspace{0.7mm} or below.      
            \item Laser-Compton backscattering: A photon beam collides with the electron beam. The photons are boosted in energy through the inverse scattering process. A quasi-monochromatic gamma beam with high intensities can be produced and used for nuclear photonics (e.g. Ref.~\cite{ERL:nucl-photonics}) or as a beam diagnostic tool (e.g., measuring beam energy width, energy). For more details, see Section~\ref{subsubsec:ERL_LCB}.
            \item Electron-ion collisions: An ion beam is collided with an electron beam as foreseen for e.g. LHeC \cite{{ERL:LHeC2012}}.
            \item Internal target experiment: The electron beam interacts with, e.g., a gas-jet target, that is directly connected to the beam pipe (e.g. MAGIX@MESA \cite{ERL:MAGIX}).
            \item Coherent electron cooling: Heavy particles are brought into thermal equilibrium with a cold electron beam. Both beams are co-propagating. In a first step the heavy particle beams' momentum distribution is modulated to the electron beam. Now, this modulation is amplified in an FEL. Finally, the density modulation in the electron beam acts on the heavy particle beam and damping its energy distribution.
        \end{itemize}

        \subsubsection{Laser-Compton Backscattering}
        \label{subsubsec:ERL_LCB}   

           Different possibilities for applications have been mentioned above. The following paragraph will give more details on the laser-Compton backscattering process.
           
\begin{figure}[h]
              \begin{center}
                \includegraphics[width=0.5\textwidth]{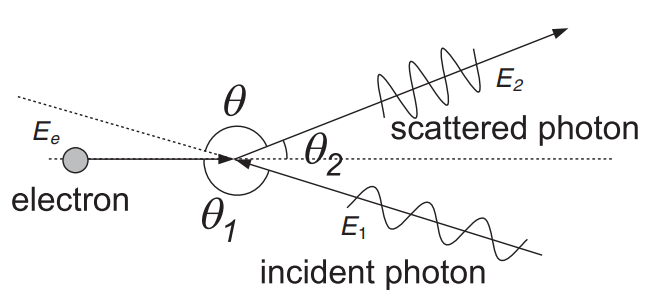}
              \end{center}
            \caption{Schematic view of the scattering process. Figure taken from \cite{ERL:LCB_paper}.}
            \label{fig:LCB_schematic}
            \end{figure}

           An electron beam is interacting with a laser beam in this collision set-up. A schematic view is shown in Fig.~\ref{fig:LCB_schematic}. An electron with energy $E_e$ collides with a photon of energy $E_1$. The photon is boosted in energy by the scattering process.

            The resulting energy $E_2$ is given by
            
            \begin{equation}
                E_2=E_1 \frac{1-\beta \cos{(\theta_1)}}{1-\beta \cos{(\theta_2)}+ (E_1/E_e)(1-\cos{(\theta)})}.
            \end{equation}   
            The electron beam is described with its velocity $\beta$. $\theta_1$ is the angle between the electron and the laser beam, $\theta_2$ between the electron beam and the scattered photons, and $\theta$ describes the angle between the~incident and scattered photons.
            
            In case of a head-on collision ($\theta_1$=0) the maximum energy of the scattered photon can be approximated by

            \begin{equation}
                E^{max}_{2}\approx 4 \gamma^2_e E_1
            \end{equation}

            \noindent for laser energies in the eV and electron energies in the higher MeV range.
            
            The flux $F=\sigma_C \cdot L$ of the generated $\gamma$-beam is also of great importance for the different experimental uses. It is given by the cross-section of the Compton scattering $\sigma_C$ and the luminosity $L$:

            \begin{equation}
                L = \frac{f N_e N_x \cos{(\phi/2)}}{2\pi \sqrt{\sigma^2_{e,y}+\sigma^2_{L,y}} \sqrt{(\sigma^2_{e,x}+\sigma^2_{L,x}) \cos^2{(\phi/2})+(\sigma^2_{e,z}+\sigma^2_{L,z}) \sin^2{(\phi/2)}}}
            \end{equation}

            \noindent with the collision frequency $f$, the number of electrons per bunch ($N_e$) respectively of photons per pulse ($N_x$), the emittances of the electron ($\sigma_{e,xyz}$) and laser beam ($\sigma_{L,xyz}$) in all three directions. The~luminosity is given for collisions at small crossing angle, $\theta_1 = \pi + \phi, \phi \ll 1$. More details can be found in \cite{ERL:LCB_paper}.

            Laser-Compton backscattering can be used as a well-suited gamma-source for e.g. Nuclear Photonics. Some requirements on the facility arise:
            \begin{itemize}
                \item A high repetition rate for a high gamma-flux by using SRF technology
                \item A high beam current at low emittance as well as recovery of the beams' kinetic energy after interaction by using an ERL
                \item A quasi-monochromatic gamma-beam by the scattering process itself
                \item A gamma-beam energy in the MeV-range by using e.g. a laser with 1 eV and a multi-turn ERL with for example around 500 MeV.
            \end{itemize}
            These requirements lead to a multi-turn SRF ERL. Having a closer look at the state-of-the-art (see Fig.~\ref{fig:overview_ERLs}), further development in the field is needed for this ideal gamma-source for Nuclear Photonics.

\section{ERLs around the World}
\label{sec:ERL_aroundWorld}

    \subsection{Overview}
    \label{subsec:ERL_overview}

\begin{figure}[h]
          \begin{center}
            \includegraphics[width=0.8\textwidth]{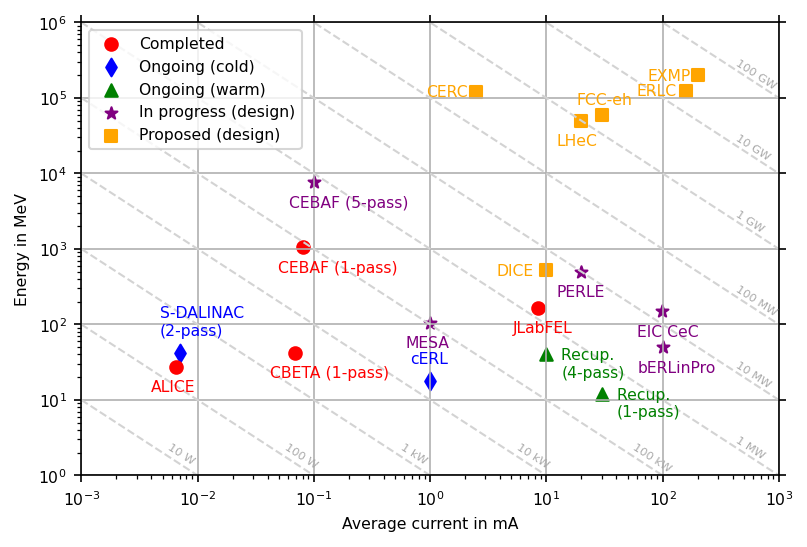}
          \end{center}
        \caption{An overview on past, present and future ERLs including their beam powers as dashed lines is given. The completed ERLs are not in operation any more. With a beam current of approx. 1~nA the 4-pass run of CBETA is outside of the scale. Ongoing projects correspond to machines being in operation at the current time. The Recup. \cite{ERL:BINP} is the only normal conducting ERL on the plot. ERLs "in progress" are under final design or construction. Proposed ERLs correspond to future projects and ideas. Figure adapted from Ref.~\cite{ERL:ERL-Roadmap}.}
        \label{fig:overview_ERLs}
        \end{figure}

        Since the idea of ERLs by Maury Tigner in 1965, many ERLs have been in operation, some are operating and many more are foreseen. Figure~\ref{fig:overview_ERLs} gives an overview of these machines including their beam energy and average current. Up to today only one ERL operated above a beam power of 1~MW (JLab FEL), three orders of magnitude less than what is planned for LHeC. Only two superconducting ERLs managed the~multi-turn mode so far (CBETA, \mbox{S-DALINAC}) that is urgently needed to achieve the anticipated beam energies. Both ERLs will be introduced in the following part of this section. To keep this chapter on the~ERL seminar compact, only two other ERLs (bERLinPro/SEALab and MESA) are presented briefly. For further information and references for the different machines, please refer to Refs.~\cite{ERL:ERL-Roadmap, ERL:ERL-review_Hutton}.

    \subsection{Example: CBETA}
    \label{subsec:ERL_CBETA}

\begin{figure}[h]
            \centering
            \begin{subfigure}{0.55\textwidth}
                \includegraphics[width=\textwidth]{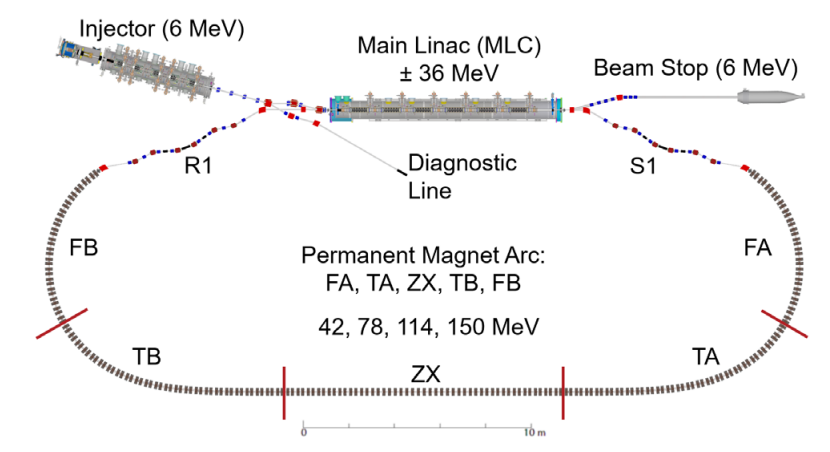}
                \caption{The layout of CBETA is given in its one-turn configuration, housing one path each in the spreader and merger section.}
                \label{fig:CBETA-1-turn-layout}
            \end{subfigure}
            \hfill
            \begin{subfigure}{0.35\textwidth}
                \includegraphics[width=\textwidth]{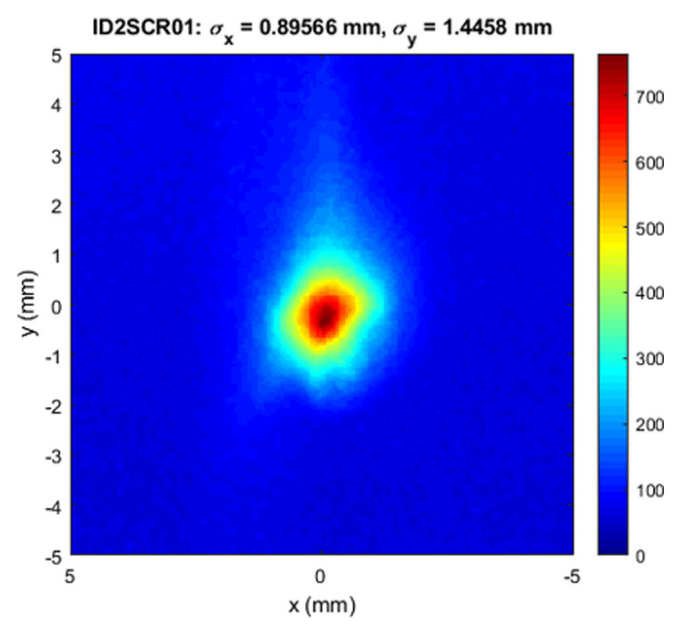}
                \caption{A beam-spot image of the beam after its deceleration at the beginning of the beam stop line.}
                \label{fig:CBETA-1-turn-beamspot}
            \end{subfigure}
            \hfill
            \begin{subfigure}{0.9\textwidth}
                \includegraphics[width=\textwidth]{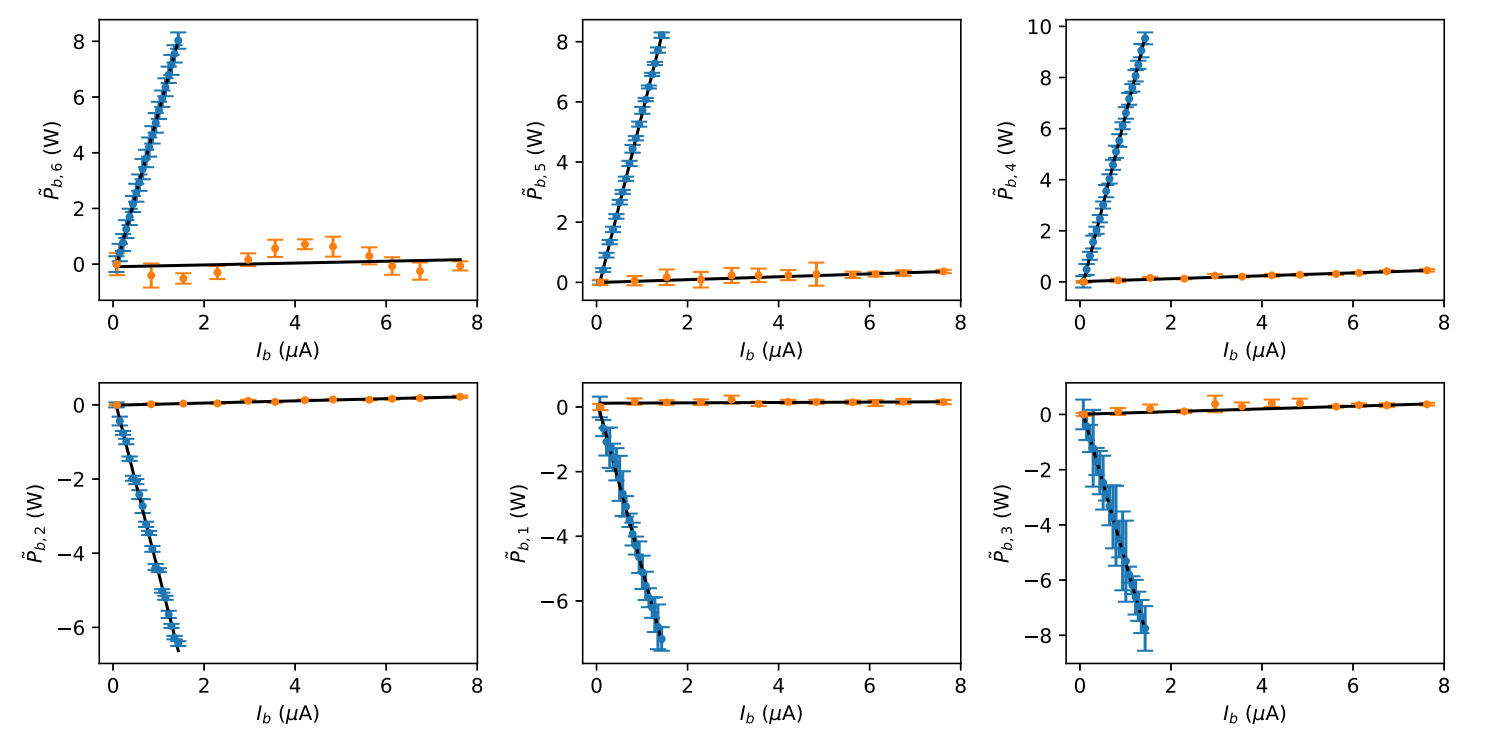}
                \caption{The beam loading of each cavity in the main linac is given as function of the beam current. The~measurement was done in two configurations: In blue three of the cavities have been used for acceleration and three for deceleration. The orange data shows the ERL configuration, where nearly no power exchange with the RF system took place.}
                \label{fig:CBETA-1-turn-ERL-eff}
            \end{subfigure}                                               
        \caption{Overview on the one-turn operation at CBETA. Figures taken from Ref.~\cite{ERL:CBETA-1-turn}.}
        \label{fig:ERL:CBETA-1-turn}
        \end{figure}

        The Cornell-BNL Test Accelerator (CBETA) is a superconducting multi-turn ERL. It was constructed and commissioned at Cornell University together with Brookhaven National Laboratory, USA. Its purpose is the R\&D for ERLs with the aim to use them in future electron-ion colliders. The lattice is composed of a superconducting injector (1.3~GHz, 6~MeV), a superconducting main linac (6 cavities, 1.3~GHz, $\pm$36~MeV), a spreader and merger section as well as a fixed-field alternating gradient (FFAG) return loop. This permanent magnet arc is capable of transporting up to seven beams of up to four different energies at the same time. The first commissioning was done in the one-turn configuration in June 2019 \cite{ERL:CBETA-1-turn}. In this set-up the spreader and merger have been constructed to allow for one acceleration and one deceleration turn (see Fig.~\ref{fig:CBETA-1-turn-layout}). During the measurement different settings have been investigated. In the ''3 up, 3 down'' configuration the first three of the main linac cavities have been used to accelerate the beam and the last three for deceleration. The measurement was performed for beam currents of up to 8~µA. The blue data in Fig.~\ref{fig:CBETA-1-turn-ERL-eff} shows the power exchange with the RF system. The one-turn ERL run data of the beam loading of all six cavities in given in the orange data. The total efficiency for this one-turn run was determined to 99.4\% \cite{ERL:CBETA-1-turn}. A beam-spot image was taken at the beginning of the beam-stop line, see Fig.~\ref{fig:CBETA-1-turn-beamspot}.

\begin{figure}[h]
            \centering
            \begin{subfigure}{0.6\textwidth}
                \includegraphics[width=\textwidth]{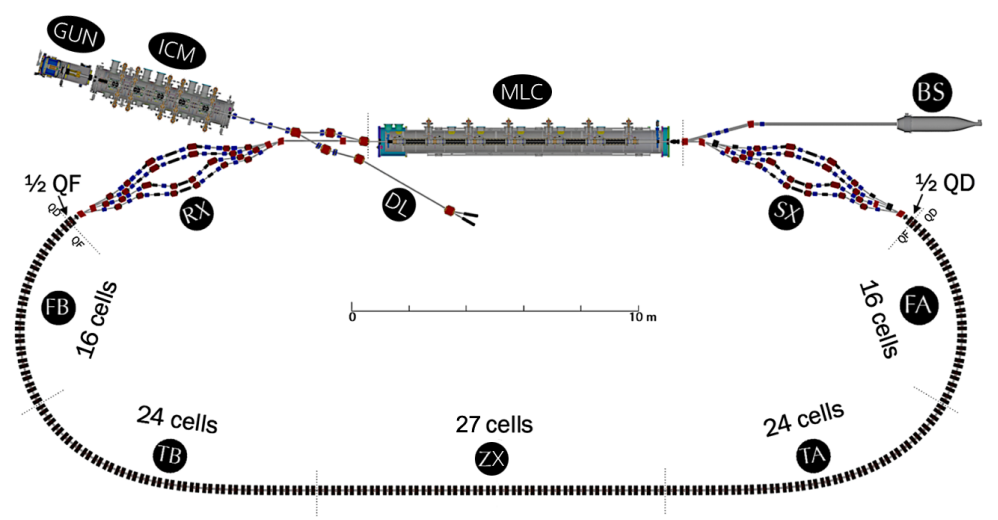}
                \caption{The layout of CBETA is given in its four-turn configuration, housing four paths each in the spreader and merger section.}
                \label{fig:CBETA-4-turn-layout}
            \end{subfigure}
            \hfill
            \begin{subfigure}{0.9\textwidth}
                \includegraphics[width=\textwidth]{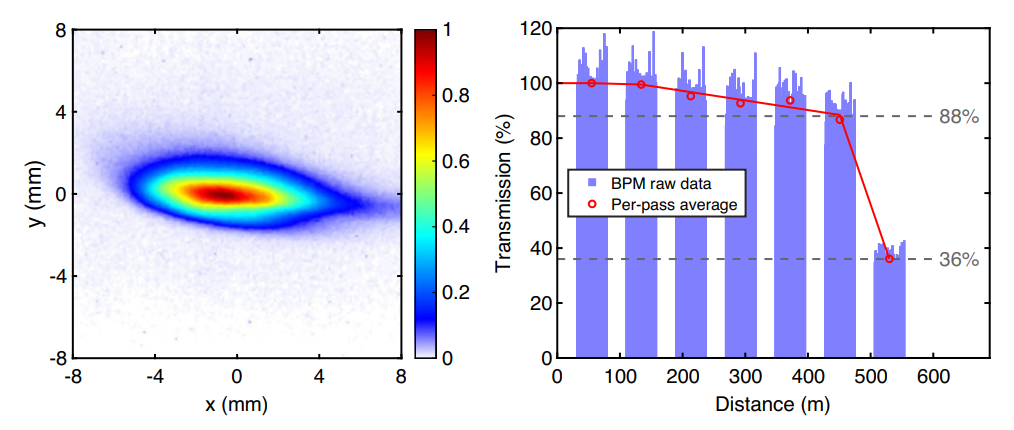}
                \caption{The left image shows a beam spot image before the beam stop. The right figure is giving an overview on the transmission for each of the seven passes through the common beam transport arc.}
                \label{fig:CBETA-4-turn-results}
            \end{subfigure}                   
         \caption{Overview on the four-turn operation at CBETA. Figures taken from Ref.~\cite{ERL:CBETA-4-turn}.}
         \label{fig:ERL:CBETA-4-turn}
         \end{figure}

        \noindent In the next step the spreader and merger sections have been re-constructed to the four-turn configuration (see Fig.~\ref{fig:CBETA-4-turn-layout}). The beam can now be accelerated and decelerated four times. This four-turn ERL operation was achieved in December 2019 \cite{ERL:CBETA-4-turn} as first operation of a superconducting multi-turn ERL world-wide, albeit at low currents of approx. 1~nA. Figure~\ref{fig:CBETA-4-turn-results} shows some of the results: An image in front of the~beam stop after the fourth deceleration passage is shown in the left part of the picture. The transmission of the~beam in each of the fixed-field return arc passages is given in the right part of this figure. A steady, but limited loss of transmission for the first six turns can be observed, followed by an additional 52\% drop in intensity for the seventh passage. The losses are the result of many small issues in the optics settings, nonlinear stray fields, microbunching and other sources. Investigations on the origins are ongoing.
      
    \subsection{Example: \mbox{S-DALINAC}}
    \label{subsec:ERL_S-DALINAC}    

\begin{figure}[h]
            \centering
            \raisebox{.3\height}
            {
                \begin{subfigure}{0.45\textwidth}
                    \includegraphics[width=\textwidth]{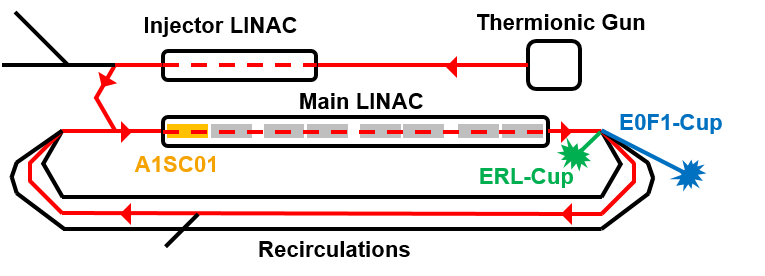}
                    \caption{The beam path for this mode is depicted in red. The RF powers of A1SC01 as well as the beam currents of the ERL-cup and E0F1-cup have been measured.}
                    \label{fig:S-DALINAC-1-turn-layout}
                \end{subfigure}
            }
            \hfill
            \begin{subfigure}{0.45\textwidth}
                \includegraphics[width=\textwidth]{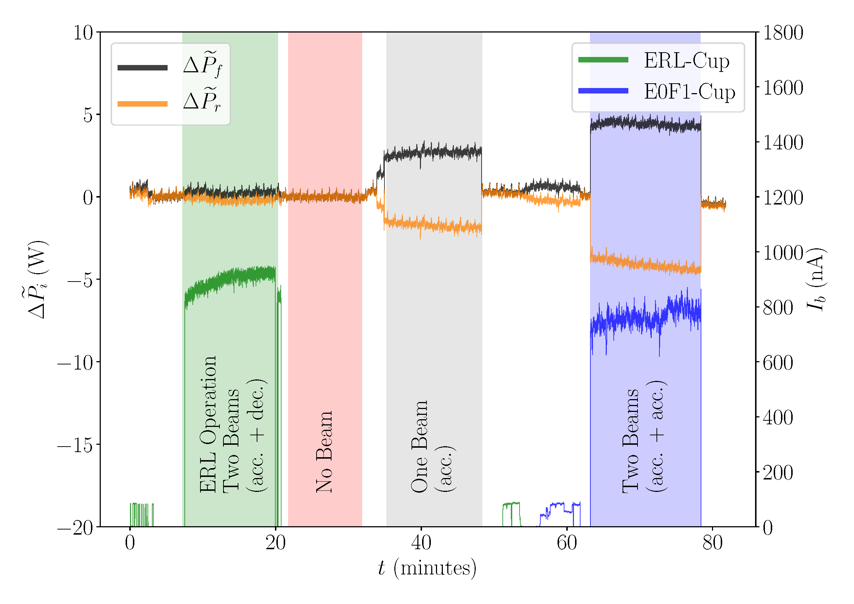}
                \caption{The RF powers (black and orange) are given during four different measurement phases: ERL mode (green), no beam in main linac (red), one accelerated beam (gray) and two accelerated beams (blue) in main linac.}
                \label{fig:S-DALINAC-1-turn-results}
            \end{subfigure}  
            \caption{Overview on the results of the one-turn ERL operation at the \mbox{S-DALINAC}. Figures taken from Ref.~\cite{ERL:S-DALINAC-1-turn}.}
            \label{fig:ERL:S-DALINAC-1-turn}
        \end{figure}

        The \mbox{S-DALINAC} is operated by Technische Universität Darmstadt, Germany since 1991 \cite{ERL:NP-IKP-paper}. It can provide electron beams of up to 130~MeV and 20~µA (recirculating) / 60~µA (injector) in cw mode at 3~GHz. It is used for research in nuclear physics and accelerator science. In 2015/2016 an upgrade of the~machine added a new recirculation beamline including the possibility for ERL operation by a~pathlength-adjustment system capable of a total phase shift of 360° \cite{ERL:S-DALINAC_Path_Length}. In 2017 a one-turn ERL mode was demonstrated \cite{ERL:S-DALINAC-1-turn}. Figure~\ref{fig:ERL:S-DALINAC-1-turn} gives an overview on this operation mode. The path of the beam is depicted in red (see Fig.~\ref{fig:S-DALINAC-1-turn-layout}). The electrons are accelerated in the injector linac (energy gain 2.5~MeV in this setting) and are bent into the main accelerator. Downstream of the main linac (energy gain of 20~MeV in this setting), the beam is recirculated in the middle recirculation with an end-energy of 22.5~MeV (in this setting). A~phase shift is applied here, so that the beam is decelerated during its next passage through the main linac. At the end the electrons are dumped at injection energy. At that time only the RF powers of the~first main linac cavity (A1SC01) could be measured. The machine was switched between four different operating modes, given by the four shaded backgrounds in Fig.~\ref{fig:S-DALINAC-1-turn-results}:

        \begin{itemize}
            \item Green: ERL operation - one accelerated and one decelerated beam in main linac (beam dumped in ERL-cup)
            \item Red: No beam in main linac, RF running in steady state
            \item Gray: One accelerated beam in main linac
            \item Blue: Two accelerated beams in main linac (beam dumped in E0F1-cup).
        \end{itemize}

        The forward and reverse RF power of A1SC01 was measured as well as the beam current on the~ERL- and E0F1-cups. The powers have been normalized to the phase with no beam in the main linac. The area between both RF-curves corresponds to the beam loading in this cavity. The beam loading for one accelerated beam (gray) is nearly doubled as a second beam is accelerated in the main linac (blue). Changing to ERL mode (green) shows nearly no beam loading left in the cavity. The measured powers are given in Table~\ref{tab:ERL-S-DALINAC-1-turn-powers}. There was no dedicated beamline optimization and thus an incomplete transmission after switching the machine to two accelerated beams. The intention was to keep the lattice setting as comparable as possible. For the ERL operation the efficiency amounts to $(90.1\pm0.3)\%$ \cite{ERL:S-DALINAC-1-turn}. The measurement was done with an initial current of 1.2~µA.
        
        \begin{table}[h]
            \begin{center}
                \caption{Measured beam loading in the first main linac cavity (A1SC01) during the different operation modes in the one-turn ERL measurement\cite{ERL:S-DALINAC-1-turn}}
                \label{tab:ERL-S-DALINAC-1-turn-powers}
                \begin{tabular}{lc}
                    \hline\hline
                    {\textbf{Operation mode}} & {\textbf{Mean Beam Loading in W}} \\
                    \hline
                    no beam (red) & $0.00\pm0.01$ \\
                    one acc. beam (gray) & $4.51\pm0.16$ \\
                    two acc. beams (blue) & $8.59\pm0.01$ \\
                    ERL (green) & $0.45\pm0.03$ \\
                    \hline\hline
                \end{tabular}
            \end{center}
        \end{table}

\begin{figure}[h]
            \centering
            \begin{subfigure}{0.5\textwidth}
                \includegraphics[width=\textwidth]{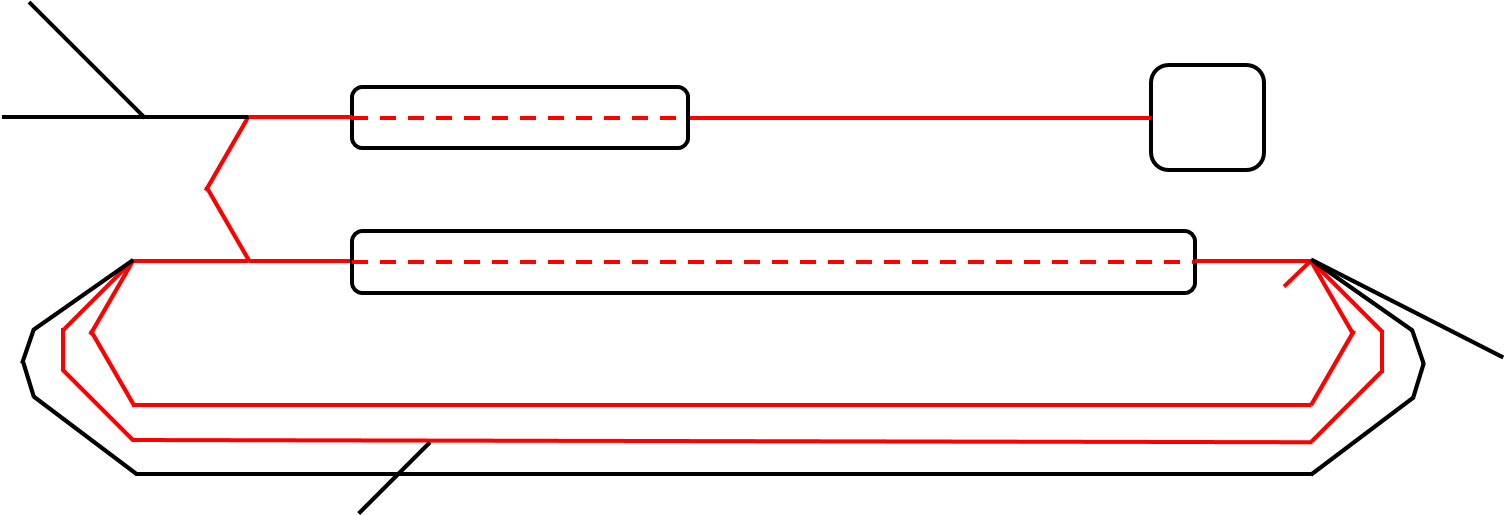}
                \caption{The beam path for this mode is depicted in red. Figure taken from \cite{ERL:S-DALINAC-1-turn}.}
                \label{fig:S-DALINAC-2-turn-layout}
            \end{subfigure}
            \\
            \hfill
            \begin{subfigure}{0.49\textwidth}
                \includegraphics[width=\textwidth]{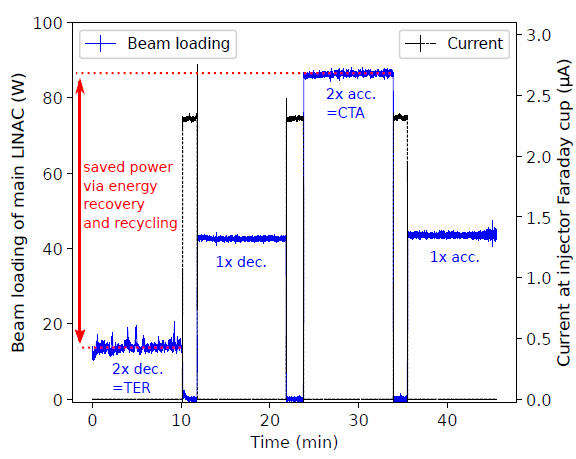}
                \caption{The total beam loading of the main accelerator is given for four different operation stages. From the right: one accelerated beam, two accelerated beams, two accelerated beams and one decelerated beam, two accelerated and two decelerated beams. Figure taken from \cite{ERL:S-DALINAC-2-turn}.}
                \label{fig:S-DALINAC-2-turn-result1}
            \end{subfigure}
            \hfill
            \raisebox{.103\height}
            {            
                \begin{subfigure}{0.49\textwidth}
                    \includegraphics[width=\textwidth]{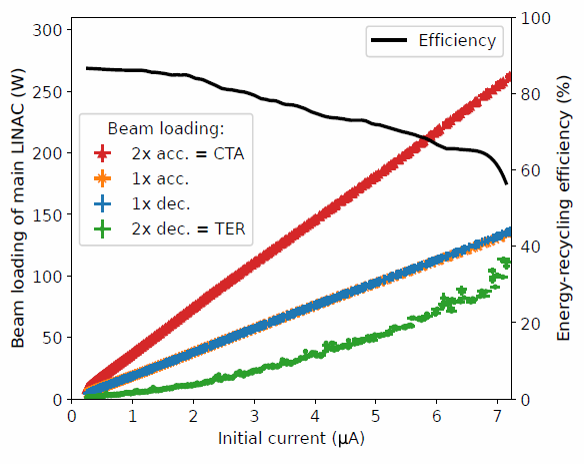}
                    \caption{The four operation stages (see Fig.~\ref{fig:S-DALINAC-2-turn-result1}) have been measured for beam currents of up to 7~µA. The energy-recycling efficiency is given. Figure taken from \cite{ERL:S-DALINAC-2-turn}.}
                    \label{fig:S-DALINAC-2-turn-result2}
                \end{subfigure}          
            }
            \caption{Overview on the results of the two-turn ERL operation at the \mbox{S-DALINAC}}
            \label{fig:ERL:S-DALINAC-2-turn}
          \end{figure}

        In 2021 the \mbox{S-DALINAC} was operated as two-turn ERL \cite{ERL:S-DALINAC-2-turn}, the first performant superconducting multi-turn ERL operation world-wide. Figure~\ref{fig:ERL:S-DALINAC-2-turn} gives an overview on the operation. The beam path is depicted in red in Fig.~\ref{fig:S-DALINAC-2-turn-layout}. The injector beam is accelerated once in the main linac and guided into the~first recirculation beamline. After a second acceleration it is transported through the second recirculation beamline including a 180° phase shift. Now the beam is decelerated for the first time, coming back to the first recirculation beamline having the same energy as before. Now two beams are travelling through this beamline with the same energy. The biggest difference is a phase shift of 180° between them, so a dedicated diagnostics system is needed. Now the beam is decelerated the second time and finally dumped at injection energy. Different phases during the two-turn ERL run have been measured similar to the one-turn ERL operation. All cavities had been equipped with RF power measurement systems in the meantime \cite{ERL:RF-power-meas}. The RF power balance of the full main accelerator was measured and summed up. The measurement was done for a fixed beam current of approx. 2.3~µA (see Fig.~\ref{fig:S-DALINAC-2-turn-result1}). Going from right to left, for each phase one beam was added to the previous setting. One accelerated beam is defining the~starting point. Adding a second accelerated beam doubled the RF power needed. In the next step the~first deceleration process took place. The total beam loading dropped back to the same level as needed to accelerate only one beam. This indicates a full recovery of the first beams' kinetic energy. In a last step the second decelerated beam was added to the machine. The RF power should drop to 0 in case of a perfect transmission. The remaining power needed to maintain this mode shows that the total efficiency is slightly below 100~\%. The same measurement was done scanning through a current range of up to 7~µA (Fig.~\ref{fig:S-DALINAC-2-turn-result2}). The data for the first decelerated beam is in perfect agreement with the power needed to accelerate one beam, showing again a nearly perfect recovery of the first beams' kinetic energy but now over a large current range. The total beam loading during the two-turn ERL operation is above 0 and steadily rising for increasing beam currents, showing losses in transmission for increasing beam currents. The calculated efficiency started at around 87~\% for small currents and dropped to below 60~\% for the~7~µA case. Most likely limitations in the transverse beam optics are the reason for the beam losses after the first deceleration. Future investigations will work on this.

\begin{figure}[h]
          \begin{center}
            \includegraphics[width=0.8\textwidth]{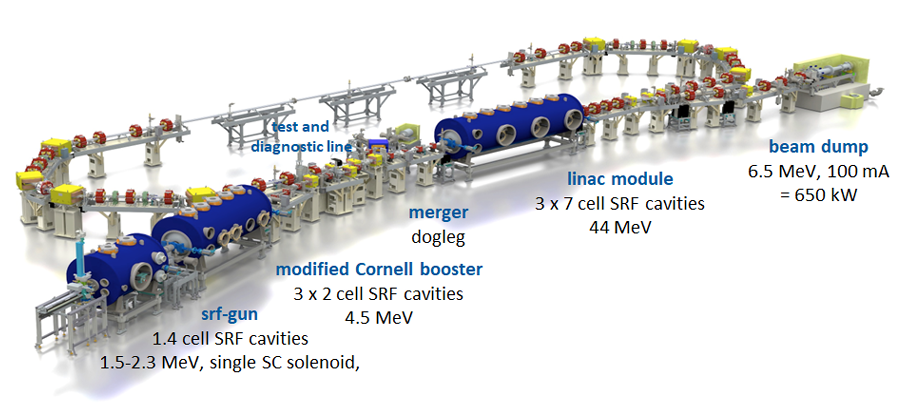}
          \end{center}
        \caption{Floorplan of bERLinPro/SEALab \cite{ERL:bERLinPro_IPAC22}}
        \label{fig:bERLinPro}
        \end{figure}

    \subsection{Overview: bERLinPro/SEALab}
    \label{subsec:ERL_bERLinPro}   

        \noindent  bERLinPro is the energy recovery linac project of Helmholtz-Zentrum Berlin, Germany. End of 2020, bERLinPro has been officially completed with the finalization of the warm part of the machine, the~building and infrastructure. The project will be continued and broadened in terms of applications. E.g., an~ultrafast electron diffraction (UED) facility producing shortest electron pulses was added to the scope of this machine. The facility is now named SEALab (Sustainable Electron Accelerator Laboratory). It is designed with a single return arc (see Fig.~\ref{fig:bERLinPro}) and can achieve a maximum beam energy of 50 MeV with a maximum average current of 100~mA. The SRF system is operated at 1.3~GHz. The parameters are summarized in Table~\ref{tab:parameters_bERLinPro}.   

        \begin{table}[h]
            \begin{center}
                \caption{Parameters of bERLinPro/SEALab \cite{ERL:bERLinPro_IPAC22}}
                \label{tab:parameters_bERLinPro}
                \begin{tabular}{lcc}
                    \hline\hline
                    {\textbf{Parameter}} & {\textbf{ERL}} & {\textbf{Injector/UED}}\\
                    \hline
                    Beam energy (MeV) & 50 & 6.5-10 / 2 \\
                    I\textsubscript{avg} (mA) & 100 & 6-10 / 0.0025 \\
                    Laser freq. (MHz) & 1300 & 50, 1300 \\
                    RF freq. (MHz) & 1300 & 1300 \\
                    $\epsilon_{\mathrm{norm}}$ (mm mrad) & 1 (0.6) & 0.6 / 0.03 \\
                    $\sigma_\mathrm{t}$ (ps) & 2 (0.1) & 0.02-2 \\
                    Bunch charge (pC) & 77 & 0.05-400 \\
                    \hline\hline
                \end{tabular}
            \end{center}
        \end{table}

        \noindent SEALab reached its commissioning phase in 2024. The first part will be the SRF photo-injector including an extensive characterization of its beam.

    \subsection{Overview: MESA}
    \label{subsec:ERL_MESA}       

\begin{figure}[h]
          \begin{center}
            \includegraphics[width=0.8\textwidth]{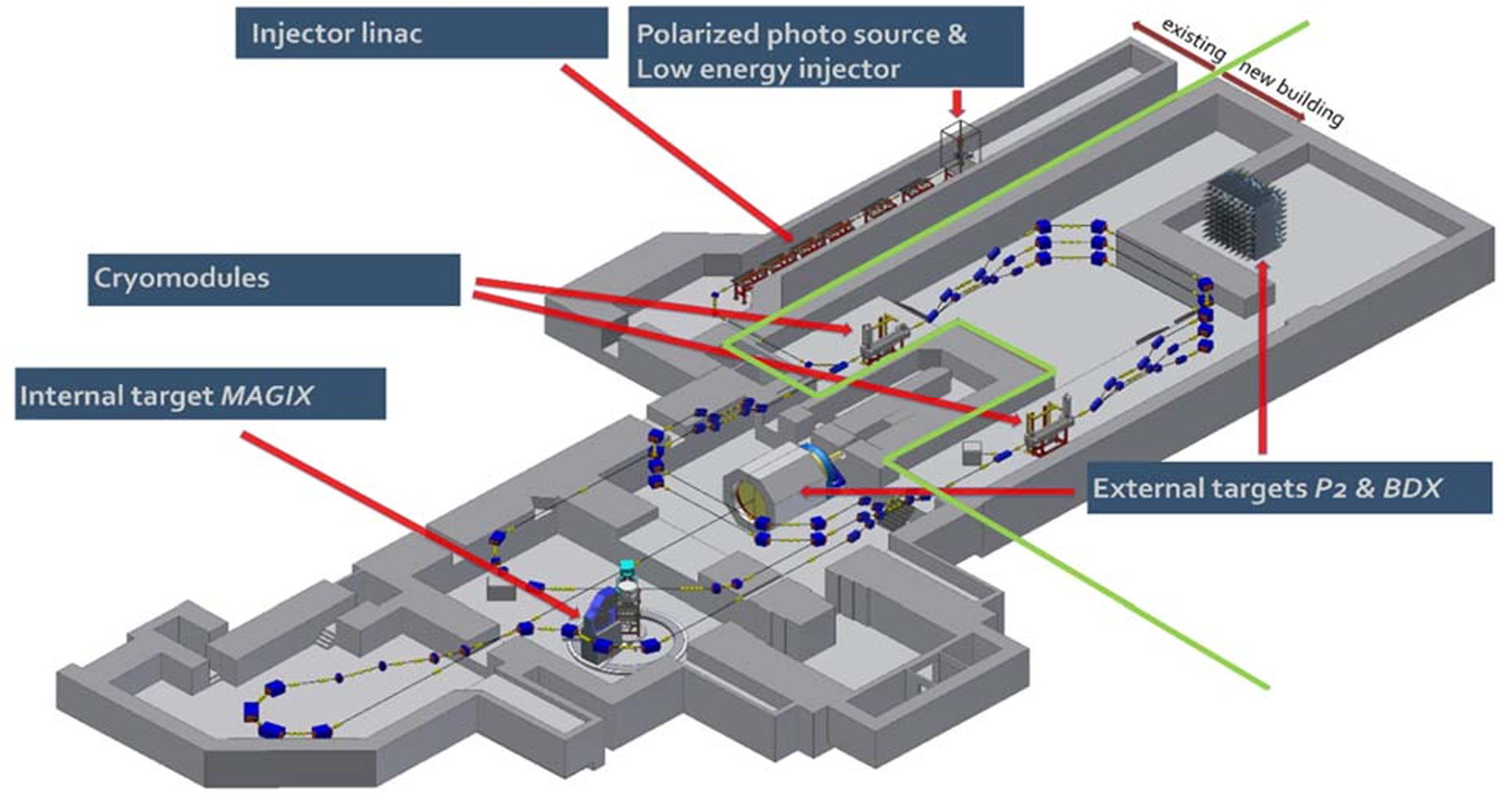}
          \end{center}
        \caption{Layout of the MESA facility \cite{ERL:MESA_ERL19}}
        \label{fig:MESA}
        \end{figure}

        MESA (Mainz Energy-Recovering Superconducting Accelerator) is an ERL under construction at Johannes Gutenberg-Universität Mainz, Germany. It is designed as a thrice-recirculating, superconducting ERL with two different operation modes: an external beam mode, where a polarized beam with up to 155~MeV and 150~$\mu$A can be provided for e.g. fixed-target experiments. In this mode the beam power will not be recovered. In the second operation mode, ERL operation with an unpolarized beam with up to 105~MeV and 1~mA will be available. In ERL mode two recirculation beamlines can be used. The~MAGIX facility will be operated in ERL scheme. Here, different experiments on gas targets can be conducted. Later the beam current in ERL-mode shall be upgraded to 10~mA (unpolarized). An overview on the facility is given in Fig.~\ref{fig:MESA}.

\section{Outlook}
\label{sec:ERL_outlook}

     ERLs represent an important accelerator design for future large-scale facilities as shown in Fig.~\ref{fig:overview_ERLs}. The~current ERLs range in the beam power from around 100~W up to the JLabFEL, the only one above 1~MW. The next-generation ERLs will cover the power range up to 10~MW, but there is still a way to go to achieve the region of 1~GW or more. A multi-turn lattice is the most promising topology for these superconducting ERLs. Only two ERLs have operated in this scheme so far: CBETA \cite{ERL:CBETA-4-turn} and \mbox{S-DALINAC} \cite{ERL:S-DALINAC-2-turn}. Both use a common beam transport scheme: at least two beams are transported in the same beamline. This concept limits the number of degrees of freedom to tune the machine. To increase the tuning possibilities and thus the chance of a high efficiency, reliability and robustness of the setting, the separated beam transport is a promising concept \cite{ERL:sep-transport}. There are two proposal under investigation, that are focusing on this transport scheme: DIANA (Daresbury Industrial Accelerator for Nuclear Applications) in the UK and DICE (Darmstadt Individually recirculating Compact ERL) in Germany (TU Darmstadt).
    
     These R\&D efforts on the multi-turn operation are complemented by the R\&D on other key aspects as low-emittance high-current sources or challenges of SRF cavities and cryompdules, see also Refs.~\cite{ERL:ERL-Roadmap,ERL:ERL-review_Hutton}. It will be a very exciting time to push all the technologies needed to enable these future machines.

\section*{Acknowledgements}
I want to thank all the persons involved in evolving this material on ERL over the years. Especially I want to thank Joachim Enders (TU Darmstadt), who gave me valuable input on the JUAS version \cite{{ERL:JUAS_Proceeding_Arnold}} of this contribution.

\begin{flushleft}
\interlinepenalty=10000

\end{flushleft}

\end{document}